\documentclass[12pt]{article} 
\usepackage[paper=letterpaper,margin=.8in]{geometry}

\usepackage{amsmath}
\usepackage{amssymb}
\usepackage{tensor}
\usepackage{graphicx}
\usepackage{caption}
\usepackage{color}
\usepackage{subfigure}
\usepackage{tikz}
\usepackage[numbers,sort&compress]{natbib}
\usepackage[utf8]{inputenc}
\usepackage{braket}
\usepackage{upgreek}
\usepackage[normalem]{ulem}
\usepackage[colorlinks = true, linkcolor = blue, citecolor = red, urlcolor  = blue, anchorcolor = blue]{hyperref}

\newcommand{\calo}{\mathcal{O}}

\newcommand{\e}{\epsilon}
\newcommand{\p}{\partial}
\newcommand{\la}{\langle}
\newcommand{\ra}{\rangle}
\newcommand{\wt}{\widetilde}

 \newcommand{\GG}{\Gamma}

\newcommand{\JJ}{{\cal J}}

\DeclareMathOperator{\tr}{Tr}

\numberwithin{equation}{section}

\begin{document}
\thispagestyle{empty}

\begin{center}

~\vspace{.8in}

{\LARGE \bf {Large p SYK from chord diagrams}}

\vspace{1in}

{\bf Baur Mukhametzhanov}

\vspace{.3in}

Department of Physics, Cornell University \\ Ithaca, NY 14853, USA

 \vspace{1in}

\end{center}

\begin{abstract}

The p-body SYK model at finite temperature exhibits submaximal chaos and contains stringy-like corrections to the dual JT gravity. It can be solved exactly in two different limits: ``large p'' SYK $1 \ll p \ll N$ and ``double-scaled'' SYK $N,p \to \infty$ with $\lambda = 2 p^2/N$ fixed. We clarify the relation between the two. Starting from the exact results in the double-scaled limit, we derive several observables in the large p limit. We compute euclidean $2n$-point correlators and out-of-time-order four-point function at long lorentzian times. To compute the correlators we find the relevant asymptototics of the ${\cal U}_{q}(su(1,1))$ 6j-symbol.

% In particular, we find that the density of states takes a very simple form up to non-perturbative corrections in $\lambda$. 

\end{abstract}

\pagebreak

{
\hypersetup{linkcolor=black}
\tableofcontents
}
\thispagestyle{empty}

%%%%%%%%%%%%%%%%%%%%%%%%%%%%%%%%%
%%%%%%%%%%%%%%%%%%%%%%%%%%%%%%%%%
%%%%%%%%%%%%%%%%%%%%%%%%%%%%%%%%%
%%%%%%%%%%%%%%%%%%%%%%%%%%%%%%%%%
%%%%%%%%%%%%%%%%%%%%%%%%%%%%%%%%%

\pagebreak
%%%%%%%%%%%%%%%%%%%%%%%%%%%%%%%
%%%%%%%%%%% New Commands %%%%%%%%%%%

\bibliographystyle{JHEP}

\pagenumbering{arabic}

\section{Introduction}

SYK model \cite{Sachdev_1993, Kitaev1, Kitaev2} has played a significant role in recent years as a solvable example of holographic duality \cite{Maldacena:2016hyu, Jensen:2016pah, Maldacena:2016upp, Engelsoy:2016xyb, Kitaev:2017awl, Kitaev:2018wpr}. At low temperatures it is approximately described by JT gravity in a nearly $AdS_2$ spacetime, e.g. see \cite{Mertens:2022irh} for a review and a more extensive list of references. In this regime the SYK model exhibits maximal chaos related to the presence of the black hole horizon \cite{Shenker:2013pqa}.

At finite temperatures there are two additional limits of parameters where the SYK model is exactly solvable. One is the ``large p'' SYK\footnote{In the literature this is often referred to as ``large q'' SYK, but in the present paper we reserve ``q'' for a different quantity.} \cite{Maldacena:2016hyu} in the limit $1 \ll p \ll N$ for a $p$-body SYK hamiltonian. The other solvable limit is the ``double-scaled'' SYK (DSSYK) \cite{Berkooz:2018jqr, Lin:2022rbf}, in the limit $N, p \to \infty$ while keeping the ratio $\lambda = 2p^2/N$ finite.

Our interest in these solvable limits is two-fold. One is that the SYK model at finite temperature exhibits submaximal chaos. Second, from the perspective of the holographic gravity dual, large p/DSSYK could be thought of as including stringy corrections to JT gravity. The two are in fact related \cite{Shenker:2014cwa}. Having a solvable model for both of these phenomena is valuable.

\vskip .3in

In the present paper we derive various observables in large p SYK as the limit $\lambda =2p^2/N \to 0$ of the exact results in DSSYK. We consider the partition function, euclidean correlators and OTOC at long lorentzian times. This was partially studied in \cite{Berkooz:2018jqr}. Our contribution is to clarify the relation and see that it agrees with existing results in the literature derived using different methods \cite{Maldacena:2016hyu, Gu:2021xaj, Choi:2023mab}.

%One of our findings 

In \cite{Berkooz:2018qkz}\footnote{ In \cite{Berkooz:2018qkz}\ the authors studied spin glass models, but the density of states turns out to be the same as in DSSYK.} it was found that the density of states in large p SYK takes a simple form\footnote{Up to an overall $s$-independent factor, see eq. \eqref{densityMain}.} 
\begin{align}\label{densityIntro}
\rho_q(s) &\approx {1\over (2\pi)^2 } \ {2\sin(\lambda s)\over \lambda}\  2\sinh(2\pi s) \ e^{-2\lambda s^2} + O(e^{4\pi(s-{\pi\over \lambda})}) \ , \qquad
E(s)  = -{2\over \lambda} \cos(\lambda s) \ . 
\end{align}
The second equation defines $s \in [0, \pi/\lambda]$ that parametrizes the energy. The energies relevant for large p SYK at finite temperature turn out to be 
\begin{align}
\lambda \to 0 \ , \qquad \lambda s \ \text{ - finite}.
\end{align}
In this regime the main difference of \eqref{densityIntro} from the Schwarzian density of states is the factor $e^{-2\lambda s^2}$, which changes the saddle in the partition function and correlators. At $\lambda \to 0$ and finite $s$ \eqref{densityIntro} reduces to the Schwarzian density of states.\footnote{Note that corrections in \eqref{densityIntro} are non-perturbative if we start from the double-scaled limit and take $\lambda \to 0$. In the full SYK model there are $1/q$ and $1/N$ corrections to the density that had been set to zero in the double-scaled limit.}

Another finding is that the Lyapunov exponent is given by 
\begin{align}\label{LyapIntro}
\lambda_L = {2\pi \over \beta} \left( 1 - {2\over \pi} \lambda s \right) \ ,
\end{align}
where $s$ is the energy parameter determined by the saddle in the partition function, see section \ref{PF}. The Lyapunov exponent in large p SYK was first derived in \cite{Maldacena:2016hyu}. In \eqref{LyapIntro} we relate it to the average energy in the system. In particular, the bound on chaos \cite{Maldacena:2015waa} is satisfied $\lambda_L \leq 2\pi/\beta$ since $s\geq 0$.

\vskip .3in

The paper is organized as follows. In section \ref{DSreview}, we review the exact results in DSSYK \cite{Berkooz:2018jqr} following the conventions of section 8 in \cite{Jafferis:2022wez}. In section \ref{PF}, we derive the density of states and the partition function in large p SYK. In section \ref{sec:2pt} we derive the euclidean two-point function. In section \ref{sec:4pt} we derive the lorentzian out-of-time-order (OTO) four-point function and euclidean $2n$-point functions.

\vskip .3in

{\bf Note:} After this work had been completed and was in preparation for publication, we learned that partially intersecting results were independently obtained in \cite{Goel:2023svz}. The current work was presented by the author at the ``Abu Dhabi Meeting on Theoretical Physics'' on January 10, 2023 \cite{AbuDhabi}.

\section{Double-scaled SYK review}

\label{DSreview}

We start with a brief review of correlation functions in DSSYK \cite{Berkooz:2018jqr}. Similar results in spin models were obtained in \cite{Berkooz:2018qkz}. The resemblance of spin models to DSSYK was also emphasized in \cite{Cotler:2016fpe}.

Our conventions for the normalization of the density of states and correlators will follow the section 8 in \cite{Jafferis:2022wez}, that are more convenient for matching with JT gravity in the low temperature limit. The SYK model has a hamiltonian
\begin{align}\label{SYKham}
H = i^{p/2} \sum_{i_1< \dots < i_p} J_{i_1\dots i_p} \psi_{i_1} \dots \psi_{i_p} \ , \qquad \la J_{i_1\dots i_p}^2 \ra = { \JJ^2 \over \lambda \left( N \atop p \right) } \ ,
\end{align}
where $\psi_i;  i = 1,\dots, N$ are Majorana fermions $\{\psi_i, \psi_j \} = 2\delta_{ij}$. Our normalization for $\JJ$ is different from \cite{Berkooz:2018jqr} by a factor of $\lambda = {2p^2 \over N}$ and instead follows \cite{Maldacena:2016hyu}. This will be more appropriate to compare with large p SYK results in the literature. Below we set $\JJ = 1$.

There are two interesting limits where exact correlators can be computed at finite temepratures
\begin{align}\label{DSSYKdef}
\text{``Double-scaled SYK'':}& \qquad N,p \to \infty , \quad  \lambda = {2p^2 \over N}  - \text{fixed}  \ .\\
\text{``Large p SYK'':}& \qquad 1\ll p \ll N \ .
\end{align}
In the second case, the limit can be taken in various ways. For example, one can take $N \to \infty$ with fixed $p$ first, and then take $p\to \infty$. This order was adopted in \cite{Maldacena:2016hyu}. Alternatively, one can take the double-scaling limit \eqref{DSSYKdef} first, and then take $\lambda \to 0$. In this paper we will take the latter approach. 

\subsection{Density of states}

 The density of states is given by\footnote{More precisely, $E(s)  = - {2 \over \sqrt{\lambda(1-e^{-\lambda})}} \cos(\lambda s)$. Our main interest will be the limit $\lambda \to 0$, where this makes no difference.}
\begin{align}
E(s) & = - {2 \over \lambda} \cos(\lambda s) \ , \qquad s \in [0, \pi/\lambda] \ , \\
\rho_q(s) & = {1\over 2\pi \Gamma_q(\pm 2i s)} \ .
\label{density}
\end{align}
The first equation is the definition of $s$ that parametrizes the energy. The energy spectrum has a finite support $E(s) \in [-2/\lambda, 2/\lambda]$. We also introduced an important parameter
\begin{align}
q=e^{-\lambda} 
\end{align}
that will appear throughout the paper. The density of states is expressed in terms of the q-gamma function $\Gamma_q(x)$ formally defined as
\begin{align}\label{qgamma}
\Gamma_q(x) =  (1-q)^{1-x} {(q;q)_\infty \over (q^x;q)_\infty} \ , \qquad (a;q)_\infty = \prod_{n=0}^\infty(1-a q^n) \ .
\end{align}
To understand this definition one first considers q-deformation of integers (``q-numbers'')
\begin{align}
[n]_q & = 1 + q +  \dots + q^{n-1} = {1-q^n \over 1-q}  \quad   \xrightarrow{q\to 1} \quad n \ .
\end{align}
Then for integer $n$ the q-gamma function is defined as the q-factorial
\begin{align}\label{qgamman}
\Gamma_q(n) &= [n-1]_q! \\ 
&= [1]_q [2]_q \dots [n-1]_q \ .
\end{align}
For integer $x$ in \eqref{qgamma} most factors in the two Pochhammer symbols cancel out and we are left with \eqref{qgamman}. Therefore \eqref{qgamma} is the analytic continuation of \eqref{qgamman} to complex $x$.

\vskip .3in

Clearly, in the limit $q\to 1$ we get the usual gamma-function $\lim_{q\to 1}\Gamma_q(x) = \Gamma(x)$. In particular, the density of states reduces to the Schwarzian density of states\footnote{Our conventions are related to \cite{Saad:2019lba} by $\rho(s) = 2 \rho_{\text there}(s)$ that could be achieved by rescaling $S_0$.} 
\begin{align}
\rho_q(s) & \approx {1\over 2\pi \Gamma(\pm 2is)} \\
& = {2s \over (2\pi)^2} \ 2\sinh(2\pi s) \ , \qquad \lambda \to 0 \ .
\label{Schdensity}
\end{align}
While the energy near the edge of the spectrum $E_0 = -2/\lambda$ is 
\begin{align}
E(s) - E_0 &= {2  \over \lambda} (1-\cos \lambda s) \\
& \approx  \lambda   s^2 \ , \qquad \lambda \to 0 \ .
\end{align}
In the second line we took the limit $\lambda \to 0$ with $s$ fixed. The overall $\lambda$ signals that we work in the low energy limit. Respectively, we work at low temperatures $\beta = {\wt \beta \over \lambda}$, such that $\beta E = \wt \beta s^2$ stays finite.\footnote{The relation to JT gravity is $\beta \JJ = {\beta_{JT} \over \e}, \lambda = 4G_N \e / \bar \phi_r$, where we restored $\JJ$.}

The partition function of DSSYK is of course
\begin{align}
Z(\beta) = \int_0^{\pi/\lambda} ds \ \rho_q(s) \ e^{-\beta E(s)} \ .
\end{align}
And our density of states \eqref{density} is normalized by\footnote{Our density of states is related to \cite{Berkooz:2018jqr} by ${1\over N_q}\rho_q(s) \ ds = {(q;q)_\infty\over 2\pi} (e^{\pm 2i \theta};q)_\infty d\theta $, where $\theta = \lambda s$.}
\begin{align}
Z(\beta = 0) = \int_0^{\pi/\lambda} ds \ \rho_q(s) & = {1\over \lambda (1-q)^2 (q;q)_\infty^3} 
 \equiv N_q \ .
\end{align}

\subsection{Correlation functions}

In \cite{Berkooz:2018jqr} the authors also considered operators
\begin{align}\label{Odef}
{\cal O} = i^{p'/2} \sum_{i_1 < \dots < i_{p'}} J_{i_1 \dots i_{p'}}^{(\cal O)} \ 
\psi_{i_1} \dots \psi_{i_{p'}} \ , 
\qquad \left\la \left( J^{(\cal O)}_{i_1\dots i_{p'}} \right)^2 \right\ra = \left( N \atop p' \right)^{-1} \ .
\end{align}
These are similar to the hamiltonian \eqref{SYKham}, but contain a different number of fermions $p'$, as well as independent gaussian random couplings $J_{i_1 \dots i_{p'}}^{(\cal O)} $. In DSSYK we work in the limit \eqref{DSSYKdef} with 
\begin{align}
p' =    p \Delta \ ,
\end{align}
where $\Delta$ is interpreted as the scaling dimension of the operator $\cal O$, which makes sense at low temperatures where the fermions have the scaling dimension $1/p$.

One might ask why we are interested in random operators \eqref{Odef}. A technical reason is that their correlation functions can be computed exactly in the double-scaling limit. Another reason is that in large p SYK the correlators of $\cal O$ will have a very similar form to those of single fermion operators, as we will see in subsequent sections.

Correlation functions of $\calo$ take the form
\begin{align}\label{nptXtoE}
\la \tr e^{-\beta_1 H} {\cal O} e^{-\beta_2 H} \calo \dots  e^{-\beta_n H} {\cal O}  \ra 
=
{1\over Z}\int_0^{\pi/\lambda} \prod_{j=1}^n \left(ds_j \ \rho_q(s_j) e^{-\beta_j E_j(s_j) } \right) \
\la \calo_{E_1E_2} \dots \calo_{E_n E_1} \ra \ .
\end{align}
The correlators in the energy basis are computed as a sum over all ``chord diagrams'' 
\begin{align}\label{chords}
\la \calo_{E_1E_2} \dots \calo_{E_n E_1} \ra \quad = \quad 
 \raisebox{-.55in}{\includegraphics[scale=0.15]{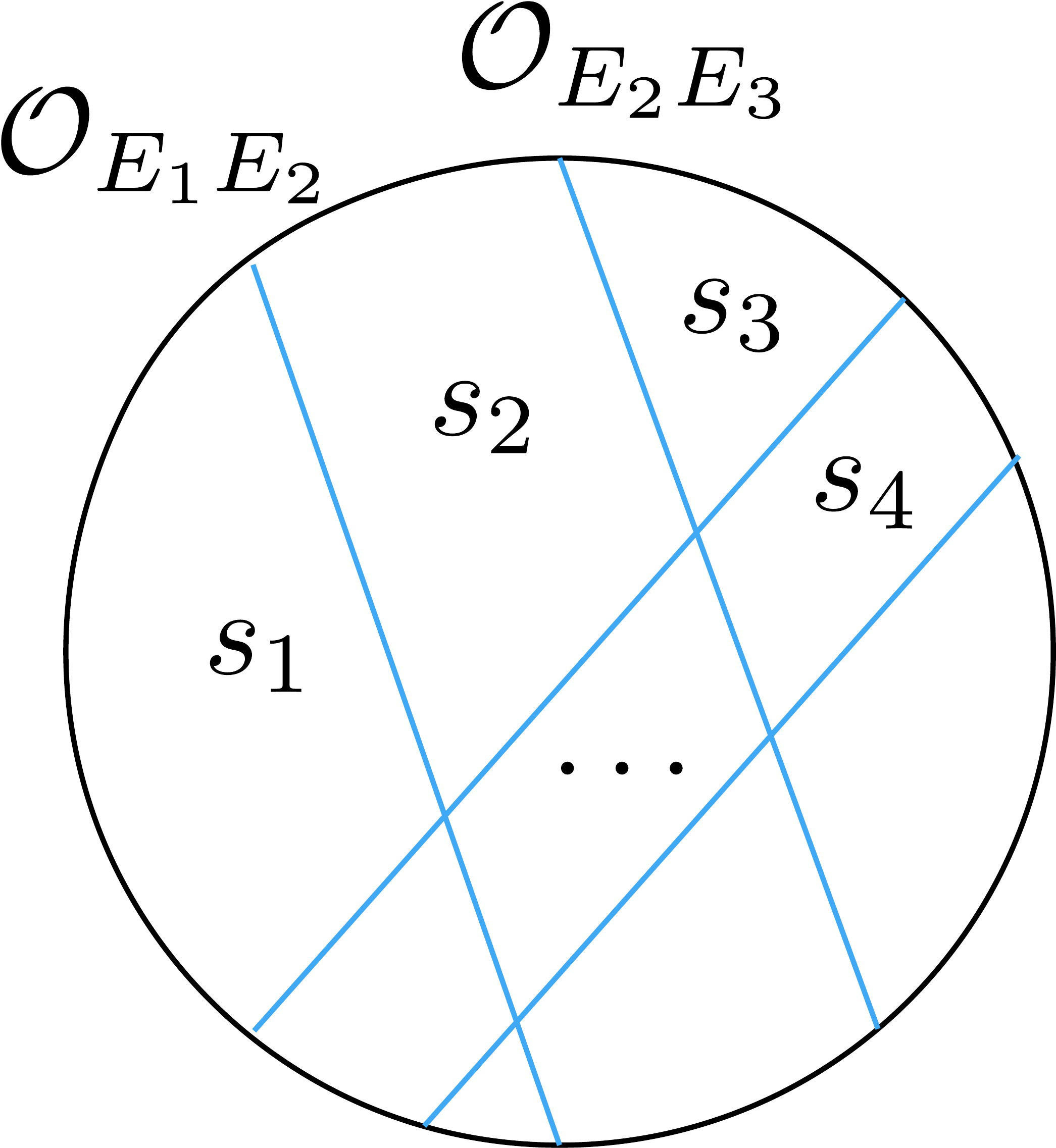}} 
 \quad + \dots
\end{align}
In each diagram operators are connected pairwise. The chords split the diagram into regions, to each of which we assign an energy parameter $s$. Then for each chord diagram we write a formula as follows:

\vskip .3in

$\bullet$ For each matrix element $O_{E_1 E_2}$ we write a factor of\footnote{We adopt a common convention that ``$\pm$'' denotes a product for all choices of signs: $\GG_q(\Delta \pm i s_1 \pm i s_2) = \GG_q(\Delta +i s_1 + is_2) \GG_q(\Delta +i s_1 - is_2) \GG_q(\Delta -i s_1 + is_2) \GG_q(\Delta -i s_1 - is_2)$}
\begin{align}
\Gamma_{12}^{1/2} \equiv \left( \GG_q(\Delta \pm i s_1 \pm i s_2) \over \GG_q(2\Delta) \right)^{1/2} \ ,
\end{align}

$\bullet$ For each intersection of chords there is a factor of the 6j-symbol of the quantum group ${\cal U}_{q^{1/2}}(su(1,1))$
\begin{align}
 \raisebox{-.35in}{\includegraphics[scale=0.2]{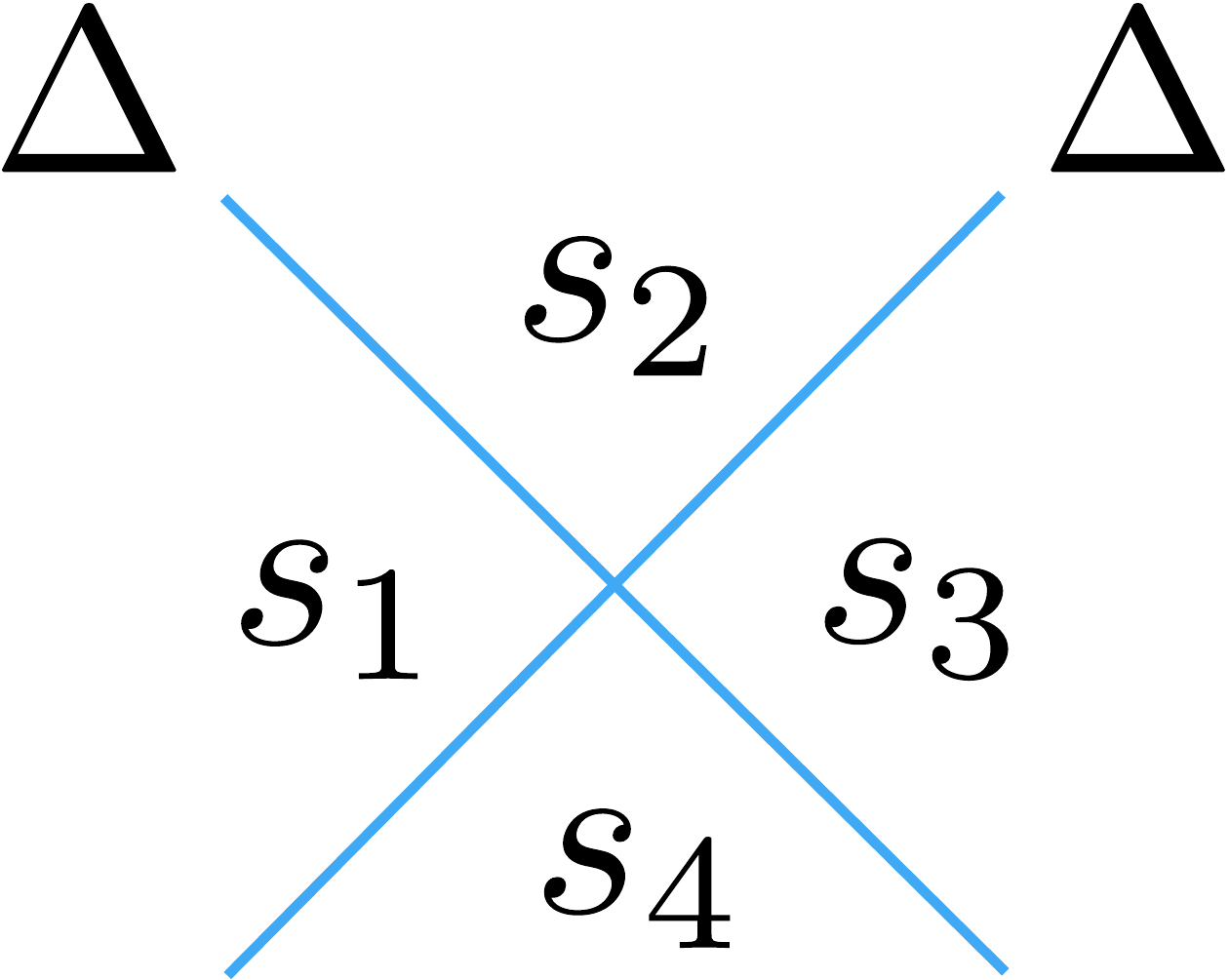}} 
 \qquad = \qquad
\left\{ 
\begin{matrix}
\Delta & s_1 & s_2 \\
\Delta & s_3 & s_4
\end{matrix}
\right\}_q \ .
\label{6jrule}
\end{align}
For now, this is just some special function with group-theoretic origin. We will discuss it in more detail in section \ref{sec:4pt}.

\vskip .3in

Let's consider some examples. The two-point function has only one diagram computed by
\begin{align}
\la \calo_{E_1E_2} \calo_{E_2E_1} \ra \quad &=  \quad
\raisebox{-.35in}{\includegraphics[scale=0.1]{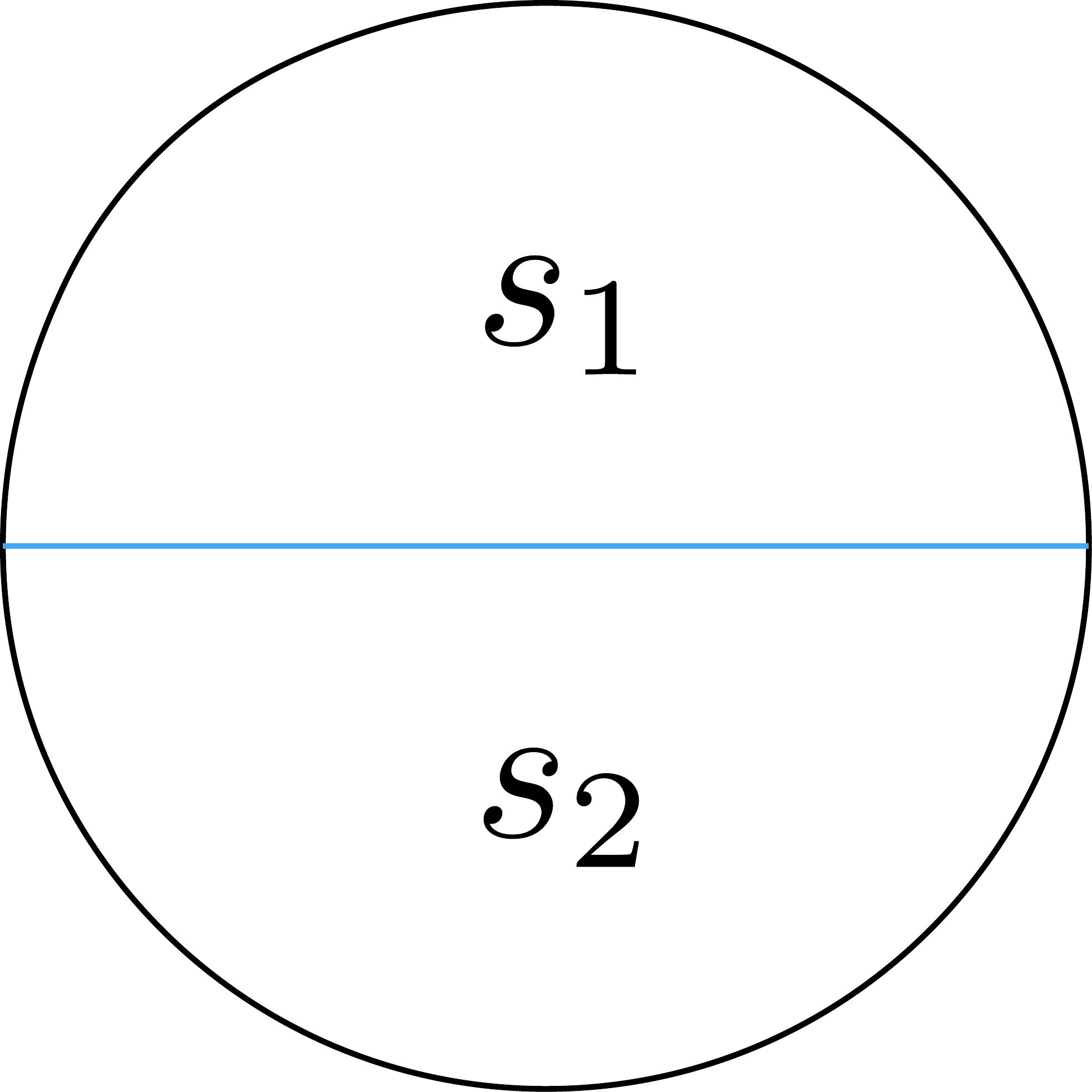}}  
\quad  =  \quad
{\GG_q(\Delta \pm i s_1 \pm i s_2) \over \GG_q(2\Delta)} \ .
\end{align}
The four-point function has three diagrams
\begin{align}\label{4ptdiag}
\la \calo_{E_1E_2} \calo_{E_2E_3} \calo_{E_3E_4} \calo_{E_4E_1} \ra 
\quad &= \quad 
\raisebox{-.3in}{\includegraphics[scale=0.08]{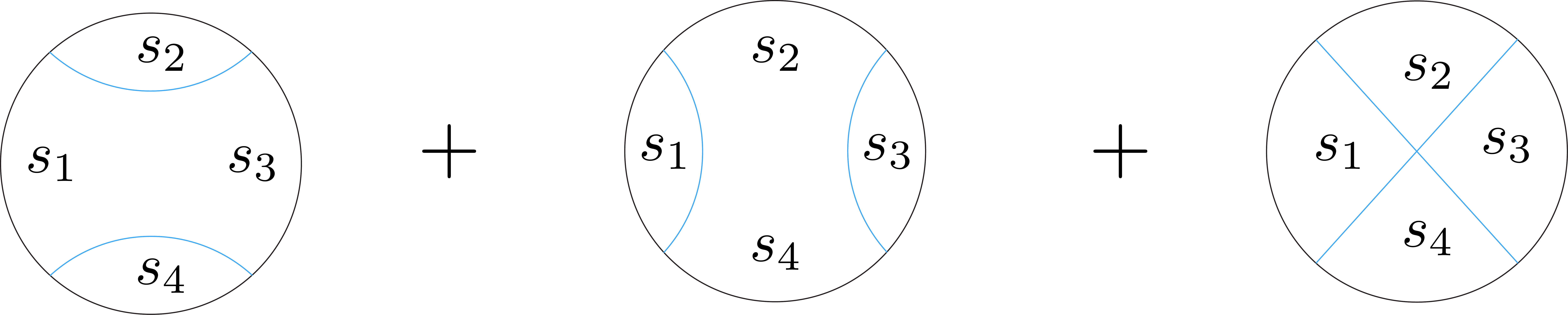}}  \\
& = 
(\GG_{12} \GG_{23}\GG_{34}\GG_{41})^{1/2}
\left({\delta(s_1 - s_3) \over \rho_q(s_1)} + {\delta(s_2 - s_4) \over \rho_q(s_2 )} +  
\left\{ 
\begin{matrix}
\Delta & s_1 & s_2 \\
\Delta & s_3 & s_4
\end{matrix}
\right\}_q
\right) \ .
\end{align}
Each region must have only one integration parameter $s$. In \eqref{nptXtoE} we introduced integrals over $s_1, \dots, s_n$ for the n-point function. Therefore, in the first two diagrams in \eqref{4ptdiag} we compensate redundant integrations by writing appropiate delta-functions.

The six-point function has $\left( 6 \atop 2 \right) = 15$ chord diagrams. Out of those, 11 are disconnected in the energy basis, i.e. products of two- and four-point functions. There are 4 connected diagrams
\begin{align}
\la \calo_{E_1E_2} \calo_{E_2E_3} \calo_{E_3E_4} \calo_{E_4E_5} \calo_{E_5E_6} \calo_{E_6E_1} \ra_{\text conn}
\quad &= \quad 
\raisebox{-.3in}{\includegraphics[scale=0.08]{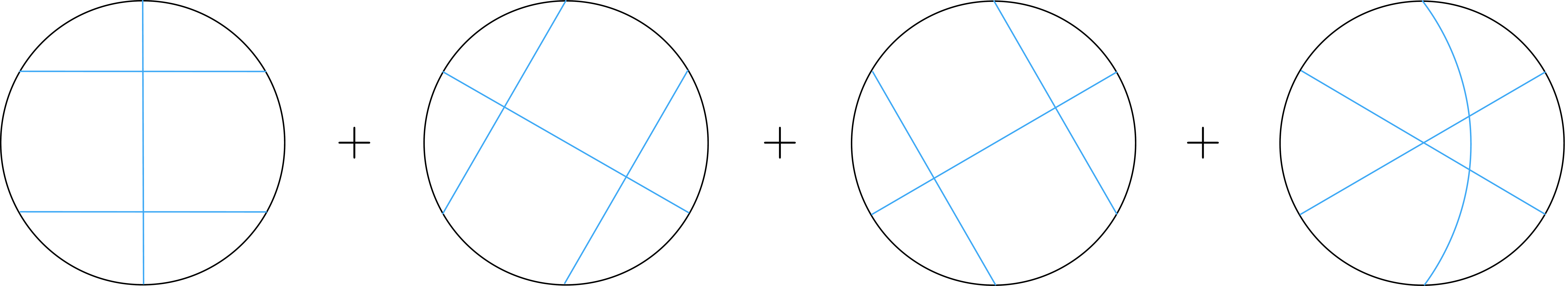}} \ . 
\end{align}

These rules for correlation functions had been first derived in \cite{Berkooz:2018jqr}.

\section{Partition function}
\label{PF}

We now turn to the computation of the partition function and correlators in large p SYK $1\ll p \ll N$. As was discussed in the previous section, this can be achieved by taking the limit $\lambda \to 0$ of DSSYK results. We are interested in keeping the temperature finite. The relevant limit for the energies will turn out to be
\begin{align}
\lambda \to 0 \ , \qquad \lambda s  \text{ - fixed} \ .
\end{align}

\subsection{Density of states}

We start with computing the relevant limit of the density of states \eqref{density}. Note that we can express it as a Jacobi function\footnote{We use conventions of \cite{polchinski_1998}.}
\begin{align}
\rho_q(s) = &
N_q  \ { \ 2\lambda \sin(\lambda s) e^{\lambda/8}  \over 2\pi} \ 
\vartheta_{11}\left(
{\lambda s \over \pi} , {i\lambda \over 2\pi} 
\right) \ ,
\end{align}
where\footnote{$(z^{\pm 1};q)_\infty \equiv (z;q)_\infty (z^{-1};q)_\infty$}
\begin{align}
\vartheta_{11}(\nu,\tau) 
=&
-{e^{\pi i \tau/4} \over 2\sin(\pi \nu)} \ (q;q)_\infty (z^{\pm 1};q)_\infty \ , \qquad
q=e^{2\pi i \tau} \ , \quad z = e^{2\pi i \nu} \ .
\end{align}
Using a modular transformation of the Jacobi function
\begin{align}
\vartheta_{11}(\nu,\tau) = {i \over \sqrt{-i\tau}} e^{-\pi i \nu^2/\tau} \vartheta_{11}(\nu/\tau,-1/\tau)
\end{align}
we find
\begin{align}
\rho_q(s) = 
  {1\over (2\pi)^2} \  {\lambda^2 \over (1-q)^2}\ {2\sin (\lambda s ) \over \lambda} \  2\sinh(2\pi s) e^{-2\lambda s^2}
 \ 
{\prod_{n=1}^\infty (1-e^{\pm 4\pi  s} q'^n) \over (q';q')_\infty^2 } \ ,
\end{align}
where $q' = e^{-4\pi^2/\lambda}$. Up to non-perturbative corrections in $\lambda$ and if we stay away from the right edge $s=\pi/\lambda$, the density of states takes a very simple form
\begin{align}\label{densityMain}
\rho_q(s) = 
{1\over (2\pi)^2} \ {\lambda^2 \over (1-q)^2}\ {2\sin (\lambda s ) \over \lambda} \  2\sinh(2\pi s) e^{-2\lambda s^2} + 
O(e^{4\pi(s-{\pi\over \lambda})}) \ .
\end{align}
If we were interested in the low temperature limit, we would take $\lambda \to 0$ with $s$ fixed. Then the above formula reduces to the Schwarzian density \eqref{Schdensity}: ${\sin(\lambda s) \over \lambda} e^{-2\lambda s^2 } \approx s$. At finite temperatures we instead keep $\lambda s$ fixed and find
\begin{align}
\rho_q(s) \approx 
{1\over (2\pi)^2}\ {2\sin (\lambda s ) \over \lambda} \  2\sinh(2\pi s) e^{-2\lambda s^2} \ .
\end{align}
It is important to retain the factors $\sin(\lambda s) e^{-2\lambda s^2}$ that will play a crucial role in computations at finite temperature and lead to a new saddle point for the average energy. These approximations were discussed in spin glass models\footnote{See eq. (4.8) in \cite{Berkooz:2018qkz}.} \cite{Berkooz:2018qkz} where the exact density of states turns out to have the same form \eqref{density}.

\subsection{Partition function}

Now we compute the partition function in the limit $\lambda \to 0$ with $\beta $ fixed
\begin{align}
Z(\beta) &= \int_0^{\pi/\lambda} ds ~ \rho_q(s) e^{-\beta E(s) } \\
&\approx  \int_0^\infty ds ~ 
 {1\over (2\pi)^2} \ {2\sin(\lambda s) \over \lambda} \ e^{2\pi s - 2\lambda s^2 +\beta {2 \over \lambda} \cos \lambda s } \ .
\end{align}
Note that if $s \sim {1\over \lambda}$ then all three terms in the exponential are large and of order $1\over \lambda$. We therefore compute in the saddle approximation
\begin{align}\label{saddle0}
\p_s \left( 
2\pi s - 2\lambda s^2 + 
 \beta {2\over \lambda} \cos \lambda s
\right) = 0
\qquad 
\Rightarrow
\quad 
2\sin(\lambda s) = {2\pi \over \beta } \left(1- {2\lambda s \over \pi} \right) \ .
\end{align}
Equivalently, the saddle-point equation is 
\begin{align}\label{saddle}
{v \over 2\cos{\pi \over 2}v } = {\beta \over 2\pi} \ , \qquad v \equiv 1- {2 \over \pi} \lambda s \ .
\end{align}
We recognize the equation derived by Maldacena and Stanford \cite{Maldacena:2016hyu}, where we also related their function $v(\beta)$ to the average energy.

The leading asymptotics of $v(\beta)$ are easy to understand from the second relation in \eqref{saddle} $v= 1- {2\over \pi} \lambda s$. Recall that $E(s) = -{2\over \lambda} \cos(\lambda s) \ , s\in [0,\pi/\lambda]$. At low temperatures the energies are near the ground state $E_0 = -2/\lambda$ and $s\approx 0$. This corresponds to $v\approx 1$. On the other hand, at high temperatures we expect to be at the maximum entropy point. This corresponds to the middle of the spectrum $s = {\pi\over 2\lambda}$ and therefore $v \approx 0$. More precisely we find from solving the saddle equation
\begin{align}
v = 
\begin{cases}
1 - {2\over \beta} + {4\over \beta^2}  \dots \ , \qquad \beta \gg 1 \ , \\
{\beta  \over \pi} + \dots \ , \qquad \beta  \ll 1\ .
\end{cases}
\end{align}

\vskip .3in

Also including gaussian fluctuations, we compute the partition function
\begin{align}\label{PartFunc}
{1\over (2\pi)^{3/2}N_q}Z(\beta)
&\approx 
{1\over 2\sqrt{\pi}  \beta^{3/2}} ~
{v \over \sqrt{\sin{\pi v \over 2}+ {2\over \beta} } } ~
\exp \left\{
{2\pi v\over \lambda}  \left(  \tan {\pi v\over 2} - {\pi v\over 4} \right)
\right\} \ .
\end{align}
The exponential precisely matches the free energy derived in \cite{Maldacena:2016hyu}. Here, we also computed the 1-loop correction that interpolates between low and high temperatures. At low temperatures $\beta  \gg 1$
\begin{align}
{1\over (2\pi)^{3/2}N_q}Z(\beta) \approx 
\left( 
1-{3\over \beta } + \dots
\right)
 {1\over 2\sqrt{\pi} \beta^{3/2}} \
 \exp\left\{
 {1\over \lambda}
 \left[
 2\beta - {\pi^2 \over 2}+ {\pi^2 \over \beta } - {2\pi^2 \over \beta^2} + \dots
 \right]
 \right\} \ .
\end{align}
In the exponent, the linear in $\beta$ term corresponds to the ground state energy $-\beta E_0 $. The second term is a correction to ground state entropy. The third term is the Schwarzian action on-shell. The fourth term is the first correction beyond the Schwarzian. At high temperatures $\beta  \ll 1$
\begin{align}
{1\over N_q}Z(\beta) \approx 
\left( 
\beta^{3/2} + \dots
\right)
{1\over \beta^{3/2}} \
\exp\left\{ 
{1\over \lambda} ( {\beta^2 \over 2} + \dots)
\right\} \ .
\end{align}
In particular, the temperature dependence cancels out in the 1-loop prefactor. We recover the correct normalization at infinite temperature ${1\over N_q} Z \approx 1$ as $\beta \to 0$.

The saddle equation \eqref{saddle0} and the partition function \eqref{PartFunc} were independently obtained in \cite{Goel:2023svz}.

\section{Two-point function}
\label{sec:2pt}

Now we compute the 2pt function in the limit $\lambda \to 0$ with $\beta$ fixed. The exact result is  \cite{Berkooz:2018jqr}
\begin{align}
\la \tr e^{-\beta H} \calo(\tau) \calo(0) \ra = 
{1\over Z} \int_0^{\pi/\lambda} ds_1 ds_2 ~ \rho_q(s_1) \rho_q(s_2) ~ e^{-\beta _1 E_1 - \beta_2 E_2} ~ {\Gamma_q(\Delta \pm is_1 \pm i s_2) \over \Gamma_q(2\Delta)} \ ,
\end{align}
where $\beta_1 = \beta - \tau, \beta_2 =  \tau$. To compute this and also other correlation functions later on, it is convenient to change variables to the average energy and energy differences. The saddle in the average energy will turn out to be the same as in the computation of the partition function \eqref{saddle}. While the integrals over energy differences will turn out to be essentially the same as in JT gravity, except for the fact that the relation between average energy and temperature is different and determined by \eqref{saddle}. For the case of the two-point function we introduce new variables 
\begin{align}
s&={1\over 2}(s_1 + s_2) \ , \qquad \omega = s_1 - s_2 \ , \\
E&= {1\over 2} (E_1 + E_2) \ , \qquad \Omega = E_1 -E_2 \ .
\end{align}
The relevant limit is $\lambda \to 0$, while keeping $\lambda s$ fixed. The differences of momenta are finite $\omega \sim 1$. In this limit the energies and densities of states are
\begin{align}
&E_j  =  -{2\over \lambda}\cos \lambda s_j  \ , \qquad (j =1,2) \ , \\
&E\approx   -{2\over \lambda}\cos \lambda s \ , \qquad 
\Omega \approx 
 \p_s E  \ \omega
 =   2\sin(\lambda s) \omega \ , \\
&\rho_q(s_1) \rho_q(s_2) \approx \rho_q(s)^2 \ .
\end{align}
We also have
\begin{align}
\beta_1 E_1 + \beta_2 E_2 &= \beta E +  \left( {\beta \over 2} - \tau \right) \Omega \\
&\approx \beta E +  \left( {\beta \over 2} - \tau \right)   \p_s E \ \omega  \ .
\label{2ptBolt}
\end{align}
Further, we can approximate the q-gamma functions as follows
\begin{align}\label{2ptG1}
\Gamma_q(\Delta \pm is_1 \pm i s_2) &= \Gamma_q(\Delta \pm 2i s ) \Gamma_q(\Delta \pm i \omega)\\
\label{2ptG2}
& = {1\over 2\pi\rho_q(s)} \ {\Gamma_q(\Delta \pm 2i s ) \over  \Gamma_q(\pm 2is)} \ 
\Gamma_q(\Delta \pm i \omega) \\
&\approx {1\over 2\pi\rho_q(s)} \left( 
2\sin(\lambda s) \over \lambda
\right)^{2\Delta}
 \ 
\Gamma(\Delta \pm i \omega) \ .
\label{2ptG3}
\end{align}
In the second line we multiplied and divided by $\Gamma_q(\pm 2is) = {1\over 2\pi \rho_q(s)}$. To compute the ratio ${\Gamma_q(\Delta \pm 2i s ) \over  \Gamma_q(\pm 2is)}$ it is convenient to use the definition as the infinite product \eqref{qgamma}. We can imagine first that $\Delta$ is an integer. Then most factors in the infinite products cancel out between $\Gamma_q(\Delta \pm 2is)$ and $\Gamma_q(\pm 2is)$. The remainder can be shown to be $\left( 
2\sin(\lambda s) \over \lambda
\right)^{2\Delta}$ in the limit $\lambda \to 0 , s \sim {1\over \lambda}$. This can of course be continued to non-integer $\Delta$. Equivalently, we use \eqref{GammaAs}. This is an analog of the formula ${\Gamma(\Delta \pm 2is) \over \Gamma(\pm 2is)} \approx (2s)^{2\Delta} , \ s \to \infty$ for the ordinary gamma function.

\vskip .3in

Putting it all together, we find
\begin{align}
{1\over Z} \int_0^\infty ds ~ \rho_q(s) e^{-\beta E(s)} \ 
\left( 
2\sin(\lambda s) \over \lambda
\right)^{2\Delta}
 \int_{-\infty}^\infty {d\omega \over 2\pi}  \ 
 e^{\omega \p_s E  (\tau - {\beta \over 2})} \ {\Gamma(\Delta \pm i \omega) \over \Gamma(2\Delta)} \ .
\end{align}
Here, the role of the integral over $s$ is to determine the relation between the average energy and temperature. This is in fact the same saddle as in the computation of the partition function \eqref{saddle}. In the rest of the integral we just set the average energy to take the saddle value. In particular, we have a factor $e^{\omega \p_s E (\tau - {\beta \over 2})}$. The saddle equation \eqref{saddle} determines $\p_s E = {2\pi v\over \beta} $. Except for this factor $\p_s E$, the integral is identical to the one arising in the semi-classical computation of JT gravity two-point function, e.g. see section 3 in \cite{Jafferis:2022uhu}. We have
\begin{align}
\left( 
{2 \cos{\pi v\over 2}\over \lambda} 
\right)^{2\Delta}
\int_{-\infty}^\infty {d\omega \over 2\pi} ~ e^{\omega {2\pi v\over \beta} (\tau - {\beta \over 2})} ~ 
{\Gamma(\Delta \pm i \omega) \over \Gamma(2\Delta)}
&=
\lambda^{-2\Delta} 
\left( 
 \cos{\pi v\over 2}
 \over 
  \cos{\pi v\over \beta}({\beta\over 2} - \tau)
\right)^{2\Delta} \ . 
\label{2ptPSYK}
\end{align}
The factor $\lambda^{-2\Delta}$ can be absorbed by rescaling the operators $\calo$. The rest agrees with the two-point function in \cite{Maldacena:2016hyu}.

\section{Four-point function}
\label{sec:4pt}

Now we turn to the computation of the four-point function in large p SYK. In euclidean signature, we will show that it factorizes into Wick contractions with the two-point function \eqref{2ptPSYK}. In fact, we will show that in euclidean signature all higher correlators factorize into products of two-point functions \eqref{2ptPSYK}. This is of course just the large N factorization and is not surprising. Our contribution is to see how this arises from chord diagrams.

More interestingly, we will compute the out-of-time-order four-point correlator (OTOC) in large p SYK. We will consider the long lorentzian time limit $u\to \infty, \lambda \to 0$ while keeping $\lambda e^{\lambda_L u}$ fixed. The Lyapunov exponent $\lambda_L$ will be given below.

To carry out these computations, the main object of interest is the 6j-symbol of the quantum group ${\cal U}_{q^{1/2}}(su(1,1))$. Many useful properties of this 6j-symbol and the related Askey-Wilson function are collected in appendix A of \cite{Jafferis:2022wez}. Other useful references include \cite{Groenevelt2, GasperRahman}. We start with writing down some explicit formulas for this object. 

\subsection{6j-symbol }

In \cite{Berkooz:2018jqr} it was computed that the crossed digram in \eqref{4ptdiag} is given in terms of a certain``basic (q-deformed) hypergeometric series" $_8\phi_7$, that we will define shortly. In \cite{Groenevelt2} it was shown that this hypergeometric series is in fact a 6j-symbol of the quantum group ${\cal U}_{q^{1/2}}(su(1,1))$. It obeys numerous identities coming from group-theoretic considerations, such as orthogonality and Yang-Baxter relations. For our purposes it is more convenient to write an equivalent representation as a sum of two $_4\phi_3$ series\footnote{See formula 8.15 in \cite{Groenevelt2}. Our definition of Askey-Wilson function differs from \cite{Groenevelt2} by $d \to 1-d$ and a factor that depends only of $q$. The former is to make our function symmetric in $a,b,c,d$. The latter is to make the limit $q\to1$ transparent. }
\begin{align}
\left\{
 \begin{matrix}
 \Delta & s_1 & s_2 \\
 \Delta & s_3 & s_4
 \end{matrix}
 \right\}_q  = &
 {\lambda \over 1-q} \ 
 \left(\gamma_{12} \gamma_{23}\gamma_{34}\gamma_{41}\right)^{1/2} 
 W_{q^{is_4}}(q^{is_2} ; q^{\Delta \pm i s_1},q^{\Delta \pm is_3}|q) \ , \qquad
\gamma_{12} \equiv \Gamma_q(\Delta \pm i s_1 \pm i s_2) \ .
\label{6jdef}
\end{align}
where the Askey-Wilson function is defined by 
\begin{align}\label{AWfunc0}
 W_{q^y}(q^{x} ; q^a,q^b,q^c,q^d|q) = &
 {\Gamma_q(d-a) \over 
 \Gamma_q(a+b) \Gamma_q(a+c) \Gamma_q(d\pm x) \Gamma_q( \wt d \pm y)
 } \\
 &_4\phi_3 \left( 
 {q^{a\pm x} , q^{\wt a \pm y} 
 \atop 
 q^{a+b} , q^{a+c} , q^{a-d+1}}
 ; q,q
 \right) + (a \leftrightarrow d) \ ,
 \label{AWfunc}
\end{align}
where the ``dual'' parameters are $\wt a = {a+b+c-d\over 2}, \wt d =  {-a+b+c+d\over 2}$. In \eqref{6jdef}, \eqref{AWfunc} ``$\pm$'' in the arguments of $W$ and $_4\phi_3$ means that they depend on both parameters and the order is not important because these functions are symmetric. The basic (q-deformed) hypergeometric series is defined as
\begin{align}
_4\phi_3 \left( 
{A_1 , A_2,A_3 , A_4
\atop 
B_1, B_2 , B_3}
; q,x
\right)
= \sum_{n=0}^\infty { (A_1, A_2, A_3, A_4;q)_n \over (B_1,B_2,B_3;q)_n } {x^n \over (q;q)_n} \ ,
\end{align}
where $(A_1,\dots, A_r;q)_n = \prod_{j=1}^r (A_j;q)_n $. 

The Askey-Wilson function \eqref{AWfunc} is symmetric in $a,b,c,d$. This implies that the 6j-symbol \eqref{6jdef} is symmetric in $s_1,s_3$. For the particular choice of parameters in \eqref{6jdef} the Askey-Wilson function, and therefore the 6j-symbol, has two additional symmetries. One is the symmetry in $s_2, s_4$ and follows from the so called ``duality'' property of the Askey-Wilson function \cite{Groenevelt2}. The other is the symmetry under the permutation of columns $s_1 \leftrightarrow s_2, \ s_3 \leftrightarrow s_4$ and follows from the representation as $_8\phi_7$, see eq. 8.14 in \cite{Groenevelt2}.

\vskip .3in

In the limit $q\to 1$ the basic hypergeometric series reduces to the hypergeometric series
\begin{align}
\lim_{q\to 1} \ 
_4\phi_3
\left( 
{q^{a_1} , q^{a_2},q^{a_3} , q^{a_4}
\atop 
q^{b_1}, q^{b_2} , q^{b_3}}
; q,x
\right) = \
_4F_3
\left( 
{a_1 , a_2, a_3 , a_4
\atop 
b_1, b_2 , b_3}
; x
\right) \ .
\end{align}
Moreover, in the limit $q\to 1$ the q-gamma functions become gamma functions. Then the Askey-Wilson function \eqref{AWfunc} becomes the Wilson function \cite{Groenevelt1, Groenevelt2}, while the quantum 6j-symbol \eqref{6jdef} becomes the classical 6j-symbol of $sl(2,{\mathbb R})$. In this way we recover the four-point function in JT gravity \cite{Mertens:2017mtv}.

\vskip .3in

Now we compute the asymptotics of the 6j-symbol. It depends on four energy parameters $s_j$. To compute the corresponding energy integrals in the four-point function \eqref{nptXtoE}, we will be interested in the limit when the average $s = {1\over 4} \sum_{j=1}^4 s_j$ is large, but the differences $s_i - s_j$ are finite. More precisely, we take
\begin{align}
\lambda \to 0 \ , \qquad \lambda s \text{ - fixed} \ , \qquad (s_i - s_j)   \text{ - fixed} \ .
\end{align}
In this limit we find (see appendix \ref{6jasympt} for details)
\begin{align}\label{6jqasympt}
\left\{
 \begin{matrix}
 \Delta & s_1 & s_2 \\
 \Delta & s_3 & s_4
 \end{matrix}
 \right\}_q \approx 
 {1\over 2\pi \rho_q(s)} \ \left( 2\sin \lambda s \over \lambda \right)^{i (\nu_2 + \nu_4)}
 \Gamma(i(\nu_2 +\nu_4)) 
 \prod_{n=1}^4 
 {\Gamma\left(\Delta - (-1)^n i \nu_n \right)^{1/2}  \over \Gamma\left(\Delta + (-1)^n i \nu_n \right)^{1/2} } + \text{c.c.} 
\end{align}
where we defined
\begin{align}\label{4ptVarChange}
s={1\over 4} \sum_{j=1}^4 s_j \ , \qquad 
\nu_1 = s_1-s_2 \ , \qquad 
\nu_2 = s_2-s_3 \ , \qquad 
\nu_3 = s_3-s_4 \ , \qquad 
\nu_4 = s_4-s_1 \ .
\end{align}
A few comments about the approximation \eqref{6jqasympt} are in order. Importantly, \eqref{6jqasympt} has large oscillations $\lambda^{-i(\nu_2 + \nu_4)}$. At long lorentzian times these oscillations will cancel with the oscillations from the long time evolution $e^{-it E_j}$, giving a finite result.

At finite euclidean times there will be no other oscillating factors in the four-point function that can cancel $\lambda^{-i(\nu_2 + \nu_4)}$. Therefore this oscillating factor effectively sets $\nu_2+\nu_4 = \nu_1 + \nu_3 = 0$. In this regime gamma functions in numerator and denominator cancel out. We also have a pole from $\Gamma(i(\nu_2 + \nu_4)) \approx {1\over i(\nu_2 + \nu_4)}$ and find in the limit $\lambda \to 0$
\begin{align}
\left\{
 \begin{matrix}
 \Delta & s_1 & s_2 \\
 \Delta & s_3 & s_4
 \end{matrix}
 \right\}_q 
 &\approx 
 {1\over 2\pi \rho_q(s)}
 \left(
 {1\over i (\nu_2 + \nu_4)}
 \left( 2\sin \lambda s \over \lambda \right)^{i (\nu_2 + \nu_4)}
  + \text{c.c.} 
 \right)
\\
&\approx {\delta(s_1 + s_3 - s_2 - s_4)\over \rho_q(s)} \ .
\label{6jqdelta}
\end{align}
In the second line we obtained a delta-function from $\lim_{\Lambda \to \infty} {2 \sin(\Lambda x) \over x} = 2\pi \delta(x)$. In our case $\Lambda = \log {2\sin(\lambda s)\over \lambda} \gg 1$.

\vskip .3in

Now we use the approximations \eqref{6jqasympt} and \eqref{6jqdelta} to compute the lorentzian OTOC and euclidean four-point functions respectively.

\subsection{OTOC at long times}

Now we compute the OTOC four-point function at long times. This comes from the crossed diagram in \eqref{4ptdiag} where we use the approximation \eqref{6jqasympt} for the 6j-symbol. We have the lorentzian correlator
\begin{align}
\la \tr e^{-\beta H} \calo(u_1) \dots \calo(u_4) \ra=
{1\over Z} \int_0^{\pi /\lambda } \prod_{j=1}^4 \left( ds_j \ \rho_q(s_j) e^{-\beta_j E_j} \right) \ 
(\Gamma_{12}\Gamma_{23}\Gamma_{34}\Gamma_{41})^{1/2} 
\left\{
 \begin{matrix}
 \Delta & s_1 & s_2 \\
 \Delta & s_3 & s_4
 \end{matrix}
 \right\}_q 
 \end{align}
 where we analytically continue to lorentzian times $u_j$ ($u_{ij} = u_i - u_j$)
 \begin{align}
& \beta_1 = \beta - iu_{14} \ , \qquad \beta_2 = iu_{12} \ , \qquad 
\beta_3 = iu_{23} \ , \qquad \beta_4 = iu_{34} \ .
\end{align}
We are interested in the OTOC regime when $u_2, u_4 \gg u_1, u_3$. It is convenient to change variables $s_1, \dots, s_4$ to the average energy and energy differences $s, \nu_1, \dots, \nu_4$ \eqref{4ptVarChange}
 \begin{align}
ds_1 ds_2 ds_3 ds_4 &= ds d\nu_1 d\nu_2 d\nu_3 \\
& = ds d\nu_1 d\nu_2 d\nu_3 d\nu_4 \ \delta(\nu_1 + \nu_2 +\nu_3 + \nu_4) \ .
\end{align}
In the second line we introduced an extra variable to make permutation symmetries manifest at the cost of a delta-function. The inverse transformation can be written as
\begin{align}\label{ston}
& s_1 = s + {2\nu_1 +\nu_2-\nu_4 \over 4} \ , \qquad 
s_2 = s + {2\nu_2 +\nu_3-\nu_1 \over 4} \ ,\\
&s_3 = s + {2\nu_3 +\nu_4-\nu_2 \over 4} \ , \qquad 
s_4 = s + {2\nu_4 +\nu_1-\nu_3 \over 4} \ .
\label{ston2}
\end{align}
The q-gamma functions in the limit $s\sim 1/\lambda$ with $\nu_j$ finite are approximated similarly to \eqref{2ptG1} - \eqref{2ptG3}
\begin{align}
(\GG_{12}\GG_{23}\GG_{34}\GG_{41})^{1/2} &\approx 
{1\over [2\pi \rho_q(s)]^2 } 
 \left( 2\sin \lambda s \over \lambda \right)^{4\Delta } 
 \prod_{n=1}^4
 \left(
 \Gamma\left( \Delta \pm  i \nu_n \right) \over \GG(2\Delta) \right)^{1/2} \ . 
\end{align}
We also use that $\prod_{j=1}^4 \rho_q(s_j) \approx \rho_q(s)^4$. Putting it all together the four-point function takes the form
 \begin{align}\label{OTOCap1}
{1\over \Gamma(2\Delta)^2} {1\over Z} \int_0^\infty ds \ \rho_q(s) e^{-\beta E(s) } 
& \left( 2\sin \lambda s \over \lambda \right)^{4\Delta } \
\int_{-\infty}^\infty {\prod_{n=1}^4 d\nu_n \over (2\pi)^3} \delta(\sum_{n=1}^4 \nu_n) \ 
e^{-\sum_{n=1}^4 \beta_n (E_n - E)} \\ 
&\left( 2\sin \lambda s \over \lambda \right)^{i (\nu_2 + \nu_4)}
 \Gamma(i(\nu_2 +\nu_4)) 
 \prod_{n=1}^4 
 \Gamma\left(\Delta - (-1)^n i \nu_n \right) \ .
 \label{OTOCap2}
\end{align}
We dropped the second complex conjugate term in \eqref{6jqasympt} because it gives an exponentially decaying in time, instead of growing, contribution. The energy differences can be expressed as $E_n - E \approx  (s_n - s) \p_s E $ and then using \eqref{ston}, \eqref{ston2} for $s_n - s$.

\vskip .3in

Let us make a few comments about \eqref{OTOCap1}, \eqref{OTOCap2}. Similarly to the two-point function, the integral over the average energy just sets the relation between average energy and temperature according to the saddle equation \eqref{saddle}. In the rest of the integral the average energy is set to its saddle value. In fact, the rest of the integral has the same form as in JT gravity \cite{Lam:2018pvp}, except for $\p_s E = {2\pi v\over \beta}$ in the Boltzmann factors. So the only difference with the computation in the Schwarzian limit, is the relation between the average energy and temperature, which is determined by the saddle of the average energy $s$ integral. The integrals over $\nu_j$ can be computed and give\footnote{In this computation, it is convenient to use the integral identity \eqref{2ptPSYK} several times.}
\begin{align}\label{OTOCfin}
{\la {\cal O}_1 {\cal O}_2{\cal O}_3{\cal O}_4 \ra_\beta
\over \la {\cal O}_1 {\cal O}_3 \ra_\beta \  \la {\cal O}_2 {\cal O}_4 \ra_\beta } 
=&
z^{-2\Delta} U(2\Delta, 1; 1/z) \ , \qquad
z=
{\lambda \over 4\lambda_L} 
{
e^{ {\lambda_L \over 2}( -i{\beta \over 2}  +u_2 + u_4 - u_1 - u_3)} 
\over 
\cosh{\lambda_L \over 2} (i{\beta \over 2} + u_{13})
\cosh{\lambda_L \over 2} (i{\beta \over 2} + u_{24})
 } \ .
\end{align}
where $U(a,1,x) = \int_0^\infty {dy \over y} e^{-x y} {y^a \over (1+y)^{a}}$ is the confluent hypergeometric function. The Lyapunov exponent is
\begin{align}\label{Lyapunov}
\lambda_L = {2\pi v \over \beta} = \p_s E = {2\pi \over \beta} \left( 1- {2\over \pi} \lambda s \right) \ .
\end{align}
For example, we can choose a symmetric configuration of opearators
\begin{align}
i u_1 &= {3\over 4}\beta - i {u \over 2} \ , \qquad
i u_2 = {1\over 2}\beta+  i {u \over 2} \ , \qquad
i u_3 = {1\over 4}\beta - i {u \over 2} \ , \qquad
i u_4 =  i {u \over 2} \ ,\qquad
z = {\lambda \over 4\lambda_L} e^{\lambda_L u} \ .
\end{align}
The result \eqref{OTOCfin} is valid in the long time limit 
\begin{align}
\lambda \to 0 \ , \qquad z \ \text{- fixed} \ . 
\end{align}
It resums an infinite power series in $z$, interpolating between early time Lyapunov growth and late time quasinormal mode decay
\begin{align}
z^{-2\Delta} U(2\Delta, 1; 1/z) \approx 
\begin{cases}
1 - 4\Delta^2 z  \ , \quad z \ll 1 \ , \\
{\log z \over \GG(2\Delta) z^{2\Delta}} \ , \quad z \gg 1 \ .
\end{cases}
\end{align}
Note that both the early time growth and the late time decay are slower than in JT gravity. 

\vskip .3in

It is also interesting to note that the last expression in \eqref{Lyapunov} for the Lyapunov exponent makes it clear that the bound on chaos \cite{Maldacena:2015waa} is satisfied
\begin{align}
\lambda_L = {2\pi \over \beta} \left( 1 - {2\over \pi} \lambda s \right) \leq {2\pi \over \beta} 
\end{align}
since $s \geq 0$. The deviation from maximality is akin to a stringy correction \cite{Shenker:2014cwa}.

\vskip .3in

Our result \eqref{OTOCfin} agrees with \cite{Gu:2021xaj} where it was derived by summing multi-ladder diagrams and with \cite{Choi:2023mab} where it was derived from the $G \Sigma$ effective action. The leading linear in $z$ term was previously computed in \cite{Streicher:2019wek, Choi:2019bmd}. The form of OTOC \eqref{OTOCfin} is similar to JT gravity \cite{Maldacena:2016upp}, except that $z$ now has more non-trivial dependence on the temperature through the Lyapunov exponent $\lambda_L(\beta)$.

\subsection{Euclidean four-point function}

We can also compute the euclidean four-point function in large p SYK as a limit of the DSSYK correlator \eqref{nptXtoE}, \eqref{4ptdiag}. Of course, the answer we expect is that at leading order it factorizes into Wick contractions with the two-point function given in \eqref{2ptPSYK}. Nevertheless, it is interesting to see how this arises from the chord diagrams. In fact, this is similar for all higher point correlators. The result is that the approximation of the 6j-symbol as a delta-function \eqref{6jqdelta} leads to factorized $n$-point correlators.

\vskip .3in

We compute the four-point function with euclidean time separations in large p SYK, i.e. in the limit $\lambda \to 0$ and $\beta_j$ finite. We can approximate the 6j-symbol by the delta-function \eqref{6jqdelta} and the four-point function \eqref{nptXtoE}, \eqref{4ptdiag}  takes the form
\begin{align}
\la \tr e^{-\beta H} \calo(\tau_1) \dots \calo(\tau_4) \ra =
{1\over Z} \int_0^{\pi /\lambda } \prod_{j=1}^4 &\left( ds_j \ \rho_q(s_j) e^{-\beta_j E_j(s_j)} \right) \ 
(\GG_{12}\GG_{23} \GG_{34} \GG_{41})^{1/2} \\
&\left( 
{\delta(s_1 - s_3) \over \rho_q(s_1)} + {\delta(s_2 - s_4) \over \rho_q(s_2 )} + 
{\delta(s_1 + s_3 - s_2 - s_4)\over \rho_q(s)}
\right) \ ,
\label{4pteucl}
\end{align}
where $(\tau_{ij} = \tau_i - \tau_j)$
\begin{align}
& \beta_1 = \beta - \tau_{14} \ , \qquad \beta_2 = \tau_{12} \ , \qquad 
\beta_3 = \tau_{23} \ , \qquad \beta_4 = \tau_{34} \ ,  \qquad
\beta > \tau_1 > \tau_2 > \tau_3 > \tau_4 > 0 \ .
\end{align}
We will focus on computing the last term coming from the 6j-symbol. The other two terms can be computed in a similar manner. This computation is essentially identical to the semi-classical limit of the four-point function in JT gravity \cite{Jafferis:2022uhu}. 

We proceed similarly to the computation of OTOC in the previous subsection. We change variables to the average energy $s$ and energy differences $\nu_1, \dots, \nu_4$ and make the same approximations for the q-gamma functions. We find for the last term in \eqref{4pteucl}
\begin{align}
  &{1\over Z} 
  \int_0^\infty ds \ \rho_q(s) e^{-\beta E(s) } 
 \left( 2\sin \lambda s \over \lambda \right)^{4\Delta }  \\
&\int_{-\infty}^\infty {d\nu_1 d\nu_2 d\nu_3 d\nu_4 \over (2\pi)^2} 
\exp\left\{ 
- \nu_1\p_s E \ {\beta_{1-2-3+4} \over 2}
- \nu_2\p_s E \ { \beta_{1+2-3-4}  \over 2}
\right\} 
\prod_{n=1}^4 \left(  \Gamma(\Delta \pm i \nu_n) \over \Gamma(2\Delta) \right)^{1/2} 
 \delta(\nu_1 + \nu_3) \delta(\nu_2+ \nu_4) \ ,
\end{align}
where $\beta = \sum_{j=1}^4 \beta_j$ and $\beta_{1-2+3-4} = \beta_1-\beta_2 +\beta_3 - \beta_4$. The integral over the average energy $s$ has the saddle \eqref{saddle}. In the rest of the integral we set the average energy to take the saddle value, e.g. $\lambda_L = \p_s E = {2\pi v\over \beta}$. Two integrals, e.g. over $\nu_3, \nu_4$, are straightforward using the delta-functions. The remaining two integrals give
\begin{align}
&\int_{-\infty}^\infty d\nu_1 d\nu_2 \ 
\exp\left\{-\nu_1\p_s E \  ({\beta\over 2} - \tau_{13})
-\nu_2 \p_s E \  ({\beta\over 2} - \tau_{24})
 \right\}
 {\Gamma(\Delta \pm i \nu_1) \Gamma(\Delta \pm i \nu_2) \over \Gamma(2\Delta)^2} 
 \\
=& 
 \lambda^{-4\Delta}
\left(
{\cos {\pi v\over 2} \over \cos{\pi v\over 2}({\beta\over 2} - \tau_{13})}
 \
  {\cos {\pi v\over 2} \over \cos{\pi v\over 2}({\beta\over 2} - \tau_{24})}
 \right)^{2\Delta} \ .
\end{align}
This is just a product of two-point functions \eqref{2ptPSYK}. A similar computation for the first two terms in \eqref{4pteucl} gives the other two Wick contractions. Altogether we have a factorized answer
\begin{align}
\la \calo_1 \calo_2 \calo_3 \calo_4 \ra_\beta = 
\la \calo_1 \calo_2 \ra \la \calo_3 \calo_4 \ra_\beta +
\la \calo_1 \calo_4 \ra \la  \calo_2 \calo_3 \ra_\beta +
\la \calo_1 \calo_3 \ra \la  \calo_2 \calo_4 \ra_\beta   
\end{align}
with the two-point functions given in \eqref{2ptPSYK}.

\subsection{Euclidean $2n$-point function}

The computation for the euclidean four-point function above is in fact straightforward to generalize to any $2n$-point function and the corresponding chord diagrams \eqref{chords}. In each chord diagram we substitute all 6j-symbols by delta-functions according to \eqref{6jqdelta}. Then the energy integrals can be computed and one obtains that the $2n$-point function is given by Wick contractions with the two-point function \eqref{2ptPSYK}. This is similar to how the correlators in JT gravity in the semi-classical limit reduce to the generalized free field \cite{Mertens:2017mtv, Jafferis:2022uhu}. Corrections in $\lambda$ can also be computed if desired.

\section*{Acknowledgements}

I am grateful to Tom Hartman, Henry Lin, Juan Maldacena, Joaquin Turiaci, Jinzhao Wang for useful discussions and Daniel Jafferis, David Kolchmeyer, Julian Sonner for collaboration on related topics. I am supported by NSF grant PHY-2014071. A part of this work was completed at KITP, Santa Barbara and supported by the National Science Foundation under Grant No. NSF PHY-1748958.

\appendix

\section{Asymptotics of the 6j-symbol}

\label{6jasympt}

To derive \eqref{6jqasympt} we do the following. The parameters of the Askey-Wilson function \eqref{AWfunc} corresponding to \eqref{6jdef} are
\begin{align}
a=d^*=\Delta+is_1, \qquad b=c^*=\Delta + is_3  \ .
\end{align}
The basic hypergeometric function in \eqref{6jdef}, 
\eqref{AWfunc} can be reduced to the ordinary hypergeometric function
\begin{align}
_4\phi_3 \left( 
 {q^{\Delta +i (s_1+s_2)} , q^{\Delta+i\nu_1} , q^{\Delta +i(s_1+s_4) } , q^{\Delta -i\nu_4} 
 \atop 
 q^{2\Delta+i(s_1+s_3)} , q^{2\Delta+i(\nu_1+\nu_2)} , q^{2is_1+1}}
 ; q,q
 \right) &\approx \
 _2\phi_1 \left( 
 {  q^{\Delta+i\nu_1}, q^{\Delta -i\nu_4} 
 \atop 
   q^{2\Delta+i(\nu_1+\nu_2)} }
 ; q,q
 \right) \\
 &\approx  \ 
 _2F_1 \left( 
 {  \Delta+i\nu_1, \Delta -i\nu_4
 \atop 
   2\Delta+i(\nu_1+\nu_2) }
 ; 1
 \right) \\
& =
 {\GG(2\Delta+i(\nu_1+\nu_2)) \GG(i(\nu_2 + \nu_4)) \over 
 \GG(\Delta+i\nu_2)\GG(\Delta - i\nu_3)} \ .
 \label{HyperAs}
\end{align}
A useful formula for computing q-gamma function asymptotics is
\begin{align}\label{GammaAs}
{\GG_q(a+is) \over \GG_q(b+is)} \approx 
\left( {1-e^{-i\lambda s} \over \lambda }\right)^{a-b} \ , \qquad \lambda \to 0, \ \lambda s \ \text{- fixed}\ .
\end{align}
Then the q-gamma functions in the prefactor in \eqref{AWfunc0} are estimated as
\begin{align}
 &{\Gamma_q(d-a) \over 
 \Gamma_q(a+b) \Gamma_q(a+c) \Gamma_q(d\pm x) \Gamma_q( \wt d \pm y)
 } \\
 =& \
2\pi \rho_q(s) \ 
 {\GG_q(2is) \GG_q(-2is) \GG_q(-2is_1)\over 
 \GG_q(2\Delta +is_{1+3})  \GG_q(\Delta - i s_{1+2})  \GG_q(\Delta - is_{1+4}) 
 } 
 \
{1\over   \GG_q(2\Delta +i\nu_{1+2})  \GG_q(\Delta - i \nu_1)  \GG_q(\Delta + i \nu_4)
}
 \\
 \approx & \ 
 2\pi \rho_q(s) \ 
\left(2\sin(\lambda s) \over \lambda \right)^{-4\Delta + i(\nu_2 + \nu_4)}
 \
{1\over   \GG_q(2\Delta +i\nu_{1+2})  \GG_q(\Delta - i \nu_1)  \GG_q(\Delta + i \nu_4)
} \ ,
\label{GammaPref}
\end{align}
where we used short notation e.g. $s_{1+2} = s_1 + s_2$. In the second line we multiplied and divided by $2\pi \rho_q(s) = {1\over \GG_q(\pm 2is)}$. In the third line we used \eqref{GammaAs}. We have from \eqref{HyperAs} and \eqref{GammaPref}
\begin{align}
W_{q^{is_4}}(q^{is_2} ; q^{\Delta \pm i s_1},q^{\Delta \pm is_3}|q) \approx 
2\pi \rho_q(s) \ 
\left(2\sin(\lambda s) \over \lambda \right)^{-4\Delta + i(\nu_2 + \nu_4)} \
{
 \GG(i(\nu_2 + \nu_4)) \over 
\prod_{n=1}^4 \GG_q(\Delta +(-1)^n \  i \nu_n) )
} \ . 
\end{align}
Similarly we find 
\begin{align}
(\gamma_{12}\gamma_{23}\gamma_{34}\gamma_{41})^{1/2} &\approx 
{1\over [2\pi \rho_q(s)]^2 } 
 \left( 2\sin \lambda s \over \lambda \right)^{4\Delta } 
 \prod_{n=1}^4
 \Gamma\left( \Delta \pm  i \nu_n \right)^{1/2} \ . 
\end{align}
Putting everything together we derive \eqref{6jqasympt}.

\bibliography{refs}

\providecommand{\href}[2]{#2}\begingroup\raggedright\begin{thebibliography}{10}

\bibitem{Sachdev_1993}
S.~Sachdev and J.~Ye, \emph{Gapless spin-fluid ground state in a random quantum
  heisenberg magnet},
  \href{https://doi.org/10.1103/physrevlett.70.3339}{\emph{Physical Review
  Letters} {\bfseries 70} (1993) 3339–3342}.

\bibitem{Kitaev1}
A.~Kitaev, \emph{A simple model of quantum holography 1},  Talk at KITP, April
  7, 2015,
  \href{{http://online.kitp.ucsb.edu/online/entangled15/kitaev/}}{{http://online.kitp.ucsb.edu/online/entangled15/kitaev/}}.

\bibitem{Kitaev2}
A.~Kitaev, \emph{A simple model of quantum holography 2},  Talk at KITP, May
  27, 2015,
  \href{{http://online.kitp.ucsb.edu/online/entangled15/kitaev2/}}{{http://online.kitp.ucsb.edu/online/entangled15/kitaev2/}}.

\bibitem{Maldacena:2016hyu}
J.~Maldacena and D.~Stanford, \emph{{Remarks on the Sachdev-Ye-Kitaev model}},
  \href{https://doi.org/10.1103/PhysRevD.94.106002}{\emph{Phys. Rev. D}
  {\bfseries 94} (2016) 106002}
  [\href{https://arxiv.org/abs/1604.07818}{{\ttfamily 1604.07818}}].

\bibitem{Jensen:2016pah}
K.~Jensen, \emph{{Chaos in AdS$_2$ Holography}},
  \href{https://doi.org/10.1103/PhysRevLett.117.111601}{\emph{Phys. Rev. Lett.}
  {\bfseries 117} (2016) 111601}
  [\href{https://arxiv.org/abs/1605.06098}{{\ttfamily 1605.06098}}].

\bibitem{Maldacena:2016upp}
J.~Maldacena, D.~Stanford and Z.~Yang, \emph{{Conformal symmetry and its
  breaking in two dimensional Nearly Anti-de-Sitter space}},
  \href{https://doi.org/10.1093/ptep/ptw124}{\emph{PTEP} {\bfseries 2016}
  (2016) 12C104} [\href{https://arxiv.org/abs/1606.01857}{{\ttfamily
  1606.01857}}].

\bibitem{Engelsoy:2016xyb}
J.~Engels\"oy, T.G.~Mertens and H.~Verlinde, \emph{{An investigation of
  AdS$_{2}$ backreaction and holography}},
  \href{https://doi.org/10.1007/JHEP07(2016)139}{\emph{JHEP} {\bfseries 07}
  (2016) 139} [\href{https://arxiv.org/abs/1606.03438}{{\ttfamily
  1606.03438}}].

\bibitem{Kitaev:2017awl}
A.~Kitaev and S.J.~Suh, \emph{{The soft mode in the Sachdev-Ye-Kitaev model and
  its gravity dual}},
  \href{https://doi.org/10.1007/JHEP05(2018)183}{\emph{JHEP} {\bfseries 05}
  (2018) 183} [\href{https://arxiv.org/abs/1711.08467}{{\ttfamily
  1711.08467}}].

\bibitem{Kitaev:2018wpr}
A.~Kitaev and S.J.~Suh, \emph{{Statistical mechanics of a two-dimensional black
  hole}}, \href{https://doi.org/10.1007/JHEP05(2019)198}{\emph{JHEP} {\bfseries
  05} (2019) 198} [\href{https://arxiv.org/abs/1808.07032}{{\ttfamily
  1808.07032}}].

\bibitem{Mertens:2022irh}
T.G.~Mertens and G.J.~Turiaci, \emph{{Solvable Models of Quantum Black Holes: A
  Review on Jackiw-Teitelboim Gravity}},
  [\href{https://arxiv.org/abs/2210.10846}{{\ttfamily 2210.10846}}].

\bibitem{Shenker:2013pqa}
S.H.~Shenker and D.~Stanford, \emph{{Black holes and the butterfly effect}},
  \href{https://doi.org/10.1007/JHEP03(2014)067}{\emph{JHEP} {\bfseries 03}
  (2014) 067} [\href{https://arxiv.org/abs/1306.0622}{{\ttfamily 1306.0622}}].

\bibitem{Berkooz:2018jqr}
M.~Berkooz, M.~Isachenkov, V.~Narovlansky and G.~Torrents, \emph{{Towards a
  full solution of the large N double-scaled SYK model}},
  \href{https://doi.org/10.1007/JHEP03(2019)079}{\emph{JHEP} {\bfseries 03}
  (2019) 079} [\href{https://arxiv.org/abs/1811.02584}{{\ttfamily
  1811.02584}}].

\bibitem{Lin:2022rbf}
H.W.~Lin, \emph{{The bulk Hilbert space of double scaled SYK}},
  \href{https://doi.org/10.1007/JHEP11(2022)060}{\emph{JHEP} {\bfseries 11}
  (2022) 060} [\href{https://arxiv.org/abs/2208.07032}{{\ttfamily
  2208.07032}}].

\bibitem{Shenker:2014cwa}
S.H.~Shenker and D.~Stanford, \emph{{Stringy effects in scrambling}},
  \href{https://doi.org/10.1007/JHEP05(2015)132}{\emph{JHEP} {\bfseries 05}
  (2015) 132} [\href{https://arxiv.org/abs/1412.6087}{{\ttfamily 1412.6087}}].

\bibitem{Gu:2021xaj}
Y.~Gu, A.~Kitaev and P.~Zhang, \emph{{A two-way approach to out-of-time-order
  correlators}}, \href{https://doi.org/10.1007/JHEP03(2022)133}{\emph{JHEP}
  {\bfseries 03} (2022) 133}
  [\href{https://arxiv.org/abs/2111.12007}{{\ttfamily 2111.12007}}].

\bibitem{Choi:2023mab}
C.~Choi, F.M.~Haehl, M.~Mezei and G.~S\'arosi, \emph{{Effective description of
  sub-maximal chaos: stringy effects for SYK scrambling}},
  [\href{https://arxiv.org/abs/2301.05698}{{\ttfamily 2301.05698}}].

\bibitem{Berkooz:2018qkz}
M.~Berkooz, P.~Narayan and J.~Simon, \emph{{Chord diagrams, exact correlators
  in spin glasses and black hole bulk reconstruction}},
  \href{https://doi.org/10.1007/JHEP08(2018)192}{\emph{JHEP} {\bfseries 08}
  (2018) 192} [\href{https://arxiv.org/abs/1806.04380}{{\ttfamily
  1806.04380}}].

\bibitem{Maldacena:2015waa}
J.~Maldacena, S.H.~Shenker and D.~Stanford, \emph{{A bound on chaos}},
  \href{https://doi.org/10.1007/JHEP08(2016)106}{\emph{JHEP} {\bfseries 08}
  (2016) 106} [\href{https://arxiv.org/abs/1503.01409}{{\ttfamily
  1503.01409}}].

\bibitem{Jafferis:2022wez}
D.L.~Jafferis, D.K.~Kolchmeyer, B.~Mukhametzhanov and J.~Sonner, \emph{{JT
  gravity with matter, generalized ETH, and Random Matrices}},
  [\href{https://arxiv.org/abs/2209.02131}{{\ttfamily 2209.02131}}].

\bibitem{Goel:2023svz}
A.~Goel, V.~Narovlansky and H.~Verlinde, \emph{{Semiclassical geometry in
  double-scaled SYK}},  [\href{https://arxiv.org/abs/2301.05732}{{\ttfamily
  2301.05732}}].

\bibitem{AbuDhabi}
B.~Mukhametzhanov, \emph{Commetns on {Double}-{Scaled} {SYK}},  Abu Dhabi
  Meeting on Theoretical Physics, January 10, 2023,
  \href{{https://nyuad.nyu.edu/en/events/2023/january/abu-dhabi-meeting-on-theoretical-physics/agenda.html}}{{https://nyuad.nyu.edu/en/events/2023/january/abu-dhabi-meeting-on-theoretical-physics/agenda.html}}.

\bibitem{Cotler:2016fpe}
J.S.~Cotler, G.~Gur-Ari, M.~Hanada, J.~Polchinski, P.~Saad, S.H.~Shenker
  et~al., \emph{{Black Holes and Random Matrices}},
  \href{https://doi.org/10.1007/JHEP05(2017)118}{\emph{JHEP} {\bfseries 05}
  (2017) 118} [\href{https://arxiv.org/abs/1611.04650}{{\ttfamily
  1611.04650}}].

\bibitem{Saad:2019lba}
P.~Saad, S.H.~Shenker and D.~Stanford, \emph{{JT gravity as a matrix
  integral}},  [\href{https://arxiv.org/abs/1903.11115}{{\ttfamily
  1903.11115}}].

\bibitem{polchinski_1998}
J.~Polchinski, \emph{String Theory}, vol.~1 of \emph{Cambridge Monographs on
  Mathematical Physics}, Cambridge University Press (1998),
  \href{https://doi.org/10.1017/CBO9780511816079}{10.1017/CBO9780511816079}.

\bibitem{Jafferis:2022uhu}
D.L.~Jafferis, D.K.~Kolchmeyer, B.~Mukhametzhanov and J.~Sonner, \emph{{Matrix
  models for eigenstate thermalization}},
  [\href{https://arxiv.org/abs/2209.02130}{{\ttfamily 2209.02130}}].

\bibitem{Groenevelt2}
W.~Groenevelt, \emph{{Wilson function transforms related to Racah
  coefficients}}, \href{https://doi.org/10.48550/arXiv.math/0501511}{\emph{Acta
  Appl. Math.} {\bfseries 91} (2006) 133}
  [\href{https://arxiv.org/abs/math/0501511}{{\ttfamily math/0501511}}].

\bibitem{GasperRahman}
G.~Gasper and M.~Rahman, \emph{Basic Hypergeometric Series}, Encyclopedia of
  Mathematics and its Applications, Cambridge University Press, 2~ed. (2004),
  \href{https://doi.org/10.1017/CBO9780511526251}{10.1017/CBO9780511526251}.

\bibitem{Groenevelt1}
W.~Groenevelt, \emph{{The Wilson function transform}},
  \href{https://doi.org/10.48550/arXiv.math/0306424}{\emph{Int. Math. Res.
  Not.} (2003) 2779} [\href{https://arxiv.org/abs/math/0306424}{{\ttfamily
  math/0306424}}].

\bibitem{Mertens:2017mtv}
T.G.~Mertens, G.J.~Turiaci and H.L.~Verlinde, \emph{{Solving the Schwarzian via
  the Conformal Bootstrap}},
  \href{https://doi.org/10.1007/JHEP08(2017)136}{\emph{JHEP} {\bfseries 08}
  (2017) 136} [\href{https://arxiv.org/abs/1705.08408}{{\ttfamily
  1705.08408}}].

\bibitem{Lam:2018pvp}
H.T.~Lam, T.G.~Mertens, G.J.~Turiaci and H.~Verlinde, \emph{{Shockwave S-matrix
  from Schwarzian Quantum Mechanics}},
  \href{https://doi.org/10.1007/JHEP11(2018)182}{\emph{JHEP} {\bfseries 11}
  (2018) 182} [\href{https://arxiv.org/abs/1804.09834}{{\ttfamily
  1804.09834}}].

\bibitem{Streicher:2019wek}
A.~Streicher, \emph{{SYK Correlators for All Energies}},
  \href{https://doi.org/10.1007/JHEP02(2020)048}{\emph{JHEP} {\bfseries 02}
  (2020) 048} [\href{https://arxiv.org/abs/1911.10171}{{\ttfamily
  1911.10171}}].

\bibitem{Choi:2019bmd}
C.~Choi, M.~Mezei and G.~S\'arosi, \emph{{Exact four point function for large
  $q$ SYK from Regge theory}},
  \href{https://doi.org/10.1007/JHEP05(2021)166}{\emph{JHEP} {\bfseries 05}
  (2021) 166} [\href{https://arxiv.org/abs/1912.00004}{{\ttfamily
  1912.00004}}].

\end{thebibliography}\endgroup

\end{document}